\documentclass{PoS}

\usepackage{amssymb,amsmath,epsfig,bm,pifont}
\usepackage{amsfonts}%
\usepackage{graphics} % <------- for LatTeX + EPS

\title{Polarized nonsinglet $\Delta q_3$ and nonsinglet
fragmentation function $D^{\pi^{+}}_{u_v}$ in the analytic \\
approach to QCD}

\ShortTitle{Polarized nonsinglet $\Delta q_3$ and nonsinglet
fragmentation function $D^{\pi^{+}}_{u_v}$}

\author{A.V. Sidorov$^a$  and  \speaker{O.P. Solovtsova}$^{\;a,b}$\\
        {$^a$}Joint Institute for Nuclear Research, 141980 Dubna, Russia\\
        {$^b$}Gomel State Technical University, 246746 Gomel, Belarus \\
        E-mail: \email{olsol@theor.jinr.ru}}
\abstract{We discuss the application of an analytic approach called
the analytic perturbation theory (APT) to the QCD analysis of DIS
data. In particular, the results of the QCD analysis of a set of
`fake' data on the polarized nonsinglet $\Delta q_3$ and the
nonsinglet fragmentation function  $D^{\pi^{+}}_{u_v}$ by using the
$Q^2$-evolution within the APT are considered.
The `fake' data are constructed based on parametrization of the
polarized PDF and nonsinglet combination of the pion fragmentation
functions.
We confirm that APT can be successfully applied to QCD analysis of
$\Delta q_3(x,Q^2)$  and $D^{\pi^{+}}_{u_v}(z,Q^2)$ and that the
inequality $\Lambda_{\rm APT} > \Lambda_{\rm PT}$ obtained
previously for the $xF_3(x)$ structure function takes place.}

\FullConference{XXII International Baldin Seminar on High Energy Physics Problems,\\
        15-20 September 2014\\
        JINR, Dubna, Russia}

\begin{document}

\section{Introduction}
We study the application of an analytic approach in QCD called the
analytic perturbation theory (APT) \cite{apt96-7} to the QCD
analysis of deep inelastic scattering (DIS) data. The question is:
how does the analytic approach work in comparison with the ordinary
perturbation theory (PT)? Continuing our previous studies on the
$F_3(x,Q^2)$ structure function data \cite{SS-F3-1,SS-F3-2}, we present the
analysis in this direction for new physical quantities: polarized
parton distribution functions (pdf's) and fragmentation functions.
We construct the so-called `fake' data for the polarized nonsinglet
combination $\Delta q_3(x,Q^2)$ and nonsinglet fragmentation
function  $D^{\pi^{+}}_{u_v}(z,Q^2)$, and compare the results of
application of the PT and APT approaches in the analysis of these
quantities.
It should be noted that the application of the APT to QCD analysis
of DIS data required a generalization of the analytic approach to
the case of non-integer power of QCD running coupling. Such a
generalization \cite{BMS05,BMS06}, for example, was applied to
analyze the $F_2(x,Q^2)$ structure function behavior at small
$x$-values \cite{Cvetic:2009kw,Shaikhatdenov:2010} and to analyze
the low energy data on nucleon spin sum rules $\Gamma_1^{p,n}(Q^2)$
\cite{PSTSK09}.

\section{Theoretical framework}

%In the framework of the analytic approach
In the leading order $(LO)$ we can write the APT nonsinglet moments
$Q^2$ evolution as follows:
\begin{equation}
{\cal{M}}^{\rm APT}(N,Q^2)
 = \frac{{\cal{A}}_{\nu}(Q^{2})}
{{\cal{A}}_{\nu}(Q_{0}^{2})} \, {\cal{M}}^{\rm APT}(N,Q^2_0) \, ,~~
{\nu(N)}={{{\gamma_{NS}^{(0),N}}/{2\beta_0}}}, ~N = 2,~3, ... \,,
\label{evol-APT}
\end{equation}
where the analytic function ${\cal{A}}_{\nu}$ is derived from the
spectral representation and corresponds to the discontinuity of the
$\nu$ power of the perturbative QCD coupling, ~$\gamma_{NS}^{(0),N}$ are the
nonsinglet  one-loop anomalous dimensions, and $\beta_0=11-2n_f/3$.

The $LO$ expression for ${\cal{A}}_{\nu}$ has rather a simple
analytic form \cite{BMS05} (see also Refs.~\cite{Cvetic:2011ym,Cvetic:2013gta})
\begin{equation}
\label{a_nu} {\cal{A}}_{\nu}(Q^2/\Lambda^2)=\left[{a}_{\rm
PT}\left({
{Q^2}/{\Lambda^2}}\right)\right]^{\nu}\,- \, {{\rm
{Li}_{1-\nu}}\left({\displaystyle
\frac{\Lambda^2}{Q^2}}\right)}/{\Gamma(\nu)} \, ,
\end{equation}
where ${a}_{\rm PT}\equiv \beta_0 \alpha_{\rm PT}/(4\pi)$ and
${\rm {Li}}_{\delta}(t)= \sum_{k=1}^{\infty}t^k/k^{\delta}$
is the polylogarithm function.
The mathematical tool for numerical calculations
of ${\cal{A}}_{\nu}$ for any $ \nu \,$ up to four-loop order is
given in Refs.~\cite{KB13,Ayala:2014pha}.
It should be stressed that values of the QCD scale parameter
$\Lambda$ are different in the PT and APT approaches. The connection
between $\Lambda_{{\rm PT}}$ and $\Lambda_{{\rm APT}}$ following
from the condition $ \left[ {a}_{\rm PT}(Q^2/\Lambda^2_{{\rm
PT}}) \right]^{\nu}$  $= {\cal{A}}_{\nu}(Q^2 / \Lambda^2_{{\rm
APT}}) $ was given in Ref.~\cite{SS-F3-1}. From the previous QCD
analysis for the $F_3(x,Q^2)$ structure function data \cite{SS-F3-2}
it was obtained that
\begin{equation}
\label{N} \Lambda_{{\rm APT}}> \Lambda_{{\rm PT}}.
\end{equation}
A similar inequality was obtained from the analysis for the
inclusive $\tau$ lepton into hadronic decays data (see, e.g.,
Refs.~\cite{Yasnov,Nesterenko:2014jfa}).

\section{Fake data construction}

\subsection{Polarized nonsinglet $\Delta q_3$}

We generate `fake' data based on the results of the phenomenological
analysis of polarized DIS data presented by
Leader--Sidorov--Stamenov (LSS'10) \cite{LSS-PDF}, where the central
values and corresponding uncertainties were presented for the
parametrisation of polarised pdf's. The kinematics region of the
generated `fake' data for the nonsinglet combination $x\Delta
q_3(x,Q^2)=[ x \Delta u(x,Q^2)+ x \Delta {\bar{u}}(x, Q^2)]- [ x
\Delta d(x,Q^2)+ x \Delta {\bar{d}}(x,Q^2)]$ corresponds
approximately to the those of the combined set of data used in
Ref.~\cite{LSS-PDF}: $0.005<x<0.7$ and $ 1~{\mbox{\rm GeV}}^2 <Q^2<
65$ GeV$^2$, $4~{\mbox{\rm GeV}}^2 <W^2$.

\subsection{Nonsinglet $D^{\pi^{+}}_{u_v}(z,Q^2)$}

In the case of the nonsinglet valence combination
$D^{\pi^{+}}_{u_v}(z,Q^2)=D^{\pi^{+}}_{u}(z,Q^2)-D^{\pi^{+}}_{\bar{u}}(z,Q^2)$
the `fake' data are generated based on the results of the LSS'14
\cite{LSS-FF} phenomenological analysis of multiplicities data of
the HERMES collaboration \cite{HERMES-FF-data}. The kinematics
region of the generated `fake' data for the nonsinglet combination
$D^{\pi^{+}}_{u_v}(z,Q^2)$ corresponds approximately to those of the
HERMES pion multiplicities \cite{HERMES-FF-data}:
 $0.2<z<0.7$ and $ 1.25~{\mbox{\rm GeV}}^2 <Q^2< 10$ GeV$^2$,
$4~{\mbox{\rm GeV}}^2 <W^2$. It should be noted that within the
kinematics region of the multiplicities data of the HERMES
collaboration analyzed in Ref.~\cite{LSS-FF}, the values of the
quantity $t=-Q^2z/x\,$ \cite{Teryaev:2002wf} are not very large:
 $\mid{t} \mid \gtrsim 4.5~{\mbox{\rm GeV}}^2$.

\section{Method of the QCD analysis}
\subsection{The PT $Q^2$ evolution}
We follow the well-known approach based on the Jacobi polynomial
expansion of structure functions. This method of solution of the
Dokshitzer-Gribov-Lipatov-Altarelli-Parisi (DGLAP) evolution
equation \cite{DGLAP} was proposed in Ref.~\cite{PS} and developed
for both unpolarized \cite{Kretal} and polarized cases \cite{LSS}.
The main formula of this method allows an approximate reconstruction
of the nonsinglet structure function through a finite number of
Mellin moments. We'll use the Jacobi method for the reconstruction
of the polarized nonsinglet $\Delta q_3(x,Q^2)$ and  nonsinglet
fragmentation function  $D^{\pi^{+}}_{u_v}(z,Q^2)$:
\begin{equation}
x\Delta q_3^{N_{max}}(x,Q^2)=x^{\alpha}(1-x)^{\beta}
%\times \\ \nonumber
\sum_{n=0}^{N_{max}} \Theta_n ^{\alpha , \beta}
(x)\sum_{j=0}^{n}c_{j}^{(n)}{(\alpha ,\beta )} M_{j+2}(Q^{2}) \,,
\label{Jacobiq3}
\end{equation}
\begin{equation}
zD^{\pi^{+} N_{max} }_{u_v}(z,Q^2)=z^{\alpha}(1-z)^{\beta}
%\times \\ \nonumber
\sum_{n=0}^{N_{max}} \Theta_n ^{\alpha , \beta}
(z)\sum_{j=0}^{n}c_{j}^{(n)}{(\alpha ,\beta )} M_{j+2}(Q^{2}) \,.
\label{JacobiD}
\end{equation}
Here $\Theta_n^{\alpha,\beta}$ are the Jacobi polynomials,
$c_j^{(n)}(\alpha,\beta)$  contain $\alpha$- and $\beta$-dependent
Euler $\Gamma$-functions where $\alpha,\beta$ are the Jacobi
polynomial parameters fixed by the minimization of the error in the
reconstruction of the function.

The perturbative renormalization group $Q^2$ evolution of moments is
well known (see, e.g., \cite{Buras:1979yt}) and in the $LO$
reads as
\begin{equation} \label{evol-PT}
~~~ M^{pQCD}_{\rm i}(N,Q^2) =\frac{~[a_{\rm PT}(Q^{2})]^{\nu}} {~[
a_{\rm PT}(Q_{0}^{2})]^{\nu}}\, M_{\rm i}(N,Q^2_0) ,~~
{\nu(N)}={{{\gamma_{NS}^{(0),N}}/{2\beta_0}}}, ~N = 2,~3, ... \,.
\end{equation}
The unknown quantity ${ M}_{\rm i}(N,Q^2_0)$ could be parameterized as the
Mellin moments of the functions $\Delta q_{3}(x,Q^2)$ or
$D^{\pi^{+}}_{u_v}(z,Q^2)$ at some point, $Q^2_0$:
\begin{equation}
{{M_{\Delta q_{3}}}}(N,Q^2_0)= \int_{0}^{1}dx{x^{N-1}}x\Delta
q_{3}(x,Q_0^2)= \int_{0}^{1}dx{x^{N-2}}Ax^{a}(1-x)^{b}(1+\gamma x)
\, , \label{Mellin-1}
\end{equation}
\begin{equation}
{{M_D}}(N,Q^2_0)= \int_{0}^{1}dz{z^{N-1}}zD^{\pi^{+}}_{u_v}(z,Q_0^2)=
\int_{0}^{1}dz{z^{N-2}}Az^{a}(1-z)^{b}(1+\gamma z) \, .
\label{Mellin-2}
\end{equation}

The parameters $A$, $a$, $b$, $\gamma$ and the scale parameter
$\Lambda_{QCD}$ are found by fitting a set  of corresponding `fake'
data on  $\Delta q_3(x,Q^2)$ or $D^{\pi^{+}}_{u_v}(z,Q^2)$,
respectively. The detailed description of the fitting procedure
could be found in Ref.~\cite{KKPS2}.

\subsection{The APT $Q^2$ evolution}
In the framework of the analytical approach in QCD the expression
for the Mellin moments evolution of the polarized nonsinglet $\Delta
q_3$ and the nonsinglet valance combination of fragmentation
functions $D^{\pi^{+}}_{u_v}$ is presented by Eq.~(\ref{evol-APT}).
Similarly to the PT case, we can represented analytical moments at
some point $Q^2_0$ in the following form:
\begin{equation}
{{{\cal{M}}_{\Delta q_{3}}}}(N,Q^2_0)=
\int_{0}^{1}dx{x^{N-1}}x\Delta q_{3}(x,Q_0^2)=
\int_{0}^{1}dx{x^{N-2}}Ax^{a}(1-x)^{b}(1+\gamma x) \, ,
\label{Mellin-3}
\end{equation}
\begin{equation}
{{{\cal{M}}_D}}(N,Q^2_0)=
\int_{0}^{1}dz{z^{N-1}}zD^{\pi^{+}}_{u_v}(z,Q^2_0)=
\int_{0}^{1}dz{z^{N-2}}Az^{a}(1-z)^{b}(1+\gamma z) \, ,
\label{Mellin-4}
\end{equation}
and expressions (\ref{Jacobiq3}) and (\ref{JacobiD}) are rewritten
as
\begin{equation}
x\Delta q_3^{N_{max}}(x,Q^2)=x^{\alpha}(1-x)^{\beta}
%\times \\ \nonumber
\sum_{n=0}^{N_{max}} \Theta_n ^{\alpha , \beta}
(x)\sum_{j=0}^{n}c_{j}^{(n)}{(\alpha ,\beta )} {\cal{M}}_{{\Delta
q_{3}}}(j+2,Q^{2}) \,, \label{Jacobiq3AT}
\end{equation}
\begin{equation}
zD^{\pi^{+} N_{max}}_{u_v}(z,Q^2)=z^{\alpha}(1-z)^{\beta}
%\times \\ \nonumber
\sum_{n=0}^{N_{max}} \Theta_n ^{\alpha , \beta}
(z)\sum_{j=0}^{n}c_{j}^{(n)}{(\alpha ,\beta )}
{\cal{M}}_{D}(j+2,Q^{2}) \,. \label{JacobiDAT}
\end{equation}

As was mentioned above, the Jacobi method was applied to the QCD
analysis in the polarized case in Ref.~\cite{LSS}.
Here we apply this method in both the PT and APT approaches for
reconstruction of the $Q^2$-evolution of polarized pdf's and
fragmentation functions.

\section{Fitting results and discussion}

\begin{figure}
\begin{minipage}{.45\textwidth}
%\vspace*{0.4cm}
\hspace*{0.4cm}
\centering
\includegraphics[width=0.87\textwidth]{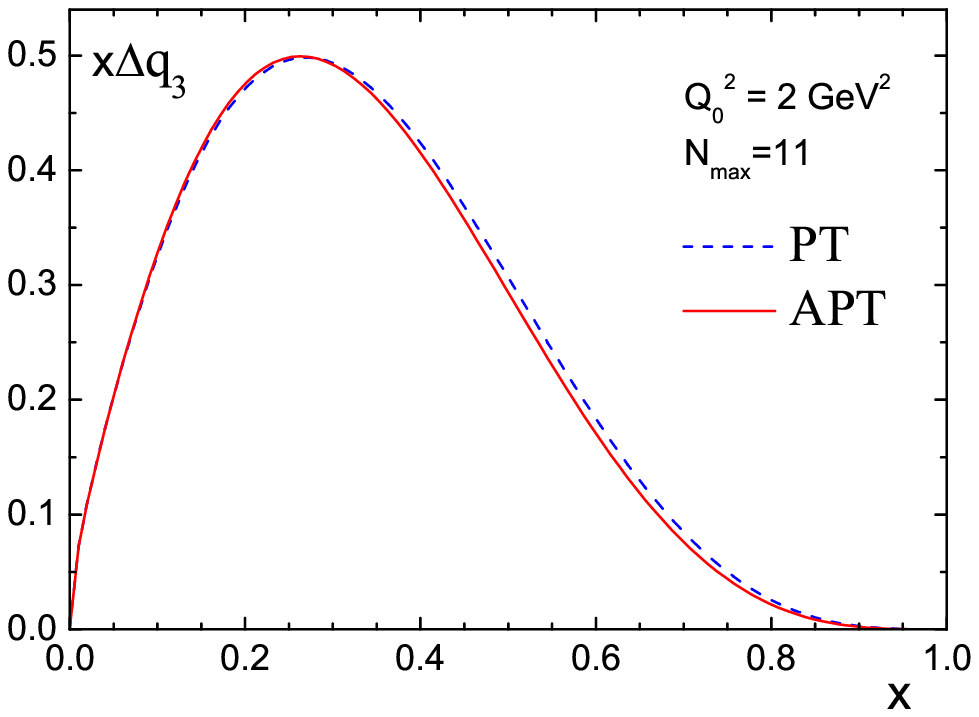}
%\vspace*{0.1cm}
\caption{ The $x{\Delta q}_3$-shape obtained in APT (solid line)
and  PT (dashed line).}
\label{figdq3}
\end{minipage}
 %\rule{.05\textwidth}{0pt}
 \phantom{}\hspace{1.0cm}%
\begin{minipage}{.45\textwidth}
%\phantom{}\vspace{0.8cm}%
\centering
\includegraphics[width=0.89\textwidth]{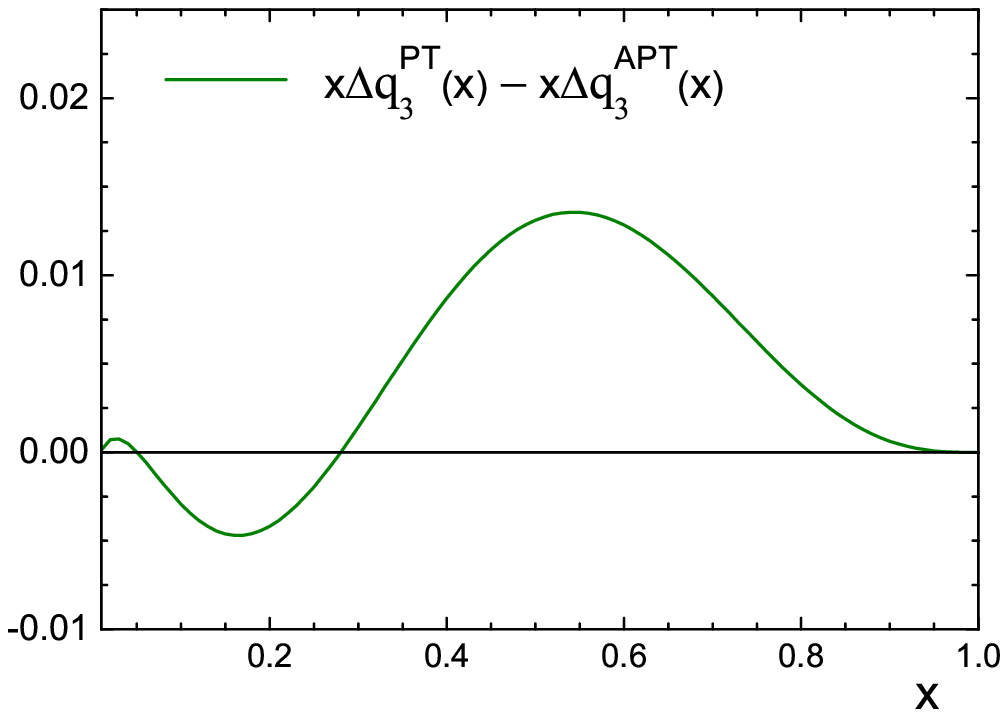}
%\vspace*{-0.97cm}
\caption{The difference in the PT and APT
for the nonsinglet combination $x{\Delta q}_3(x)$.}
\label{figdq32}
\end{minipage}
\end{figure}

The results of the $LO$ QCD fit of the `fake'  $\Delta q_{3}$ data
in the PT and APT approaches are presented in Table~1 and
Figs.~\ref{figdq3} and \ref{figdq32}. In both cases for the PT and
APT, we put $Q^2_0=2$~GeV$^2$, number of active flavors $n_f=4$ and
$N_{max}=11$. The value of errors of parameters correspond to
$\Delta{\chi}^2=1$. One can be seen from Table~1 that values of the
scale parameter $\Lambda$  are different in the PT and APT
approaches and that $\Lambda_{\rm APT} > \Lambda_{\rm PT}$.

%%%%%%%%%%%%%%%%%%%  TABLE -1 %%%%%%%%%%%%%%%%%%%%%%%%%%%%%%%%%%%%
\begin{table}[!t] \small\small
 \caption{The results for the QCD leading order
fit of the `fake'  $\Delta q_{3}$ data in the standard PT and the
APT approaches at $Q^2_0=2$~GeV$^2$, $Q^2 > 1$~GeV$^2$, $n_f=4$, and
$N_{max}=11$. }
\begin{center}\label{tab:2}
 {\begin{tabular}{|l|c|c|}   \hline
  & PT            & APT \\ [1mm]  \hline  \hline
~~ ~~ A        &  ~~~$0.807 \pm 0.091  $~~~   & ~~~$0.684 \pm 0.052 $~~~  \\
~~ ~~ $\alpha$   &  ~~~$0.536 \pm 0.024$~~~   &  ~~~$0.505 \pm 0.016 $~~~   \\
~~ ~~ $\beta$    &  ~~~~$3.43 \pm 0.023$~~~   &  ~~~~ $3.56 \pm 0.020$~~~   \\
~~ ~~ $\gamma$   &  ~~~$9.89 \pm 1.12$~~~   &  ~~~$12.55 \pm
0.87$~~~   \\ \hline ~~ ~$\Lambda$~[MeV]  &   ~~~$256 \pm11$~~   &
~~~$280 \pm15$~~~   \\ \hline \hline
\end{tabular} } \end{center}
\end{table}

Figure \ref{figdq3} shows the $x\Delta q_{3}(x)$-shape obtained in
the APT (solid line) and the PT (dotted line) cases. One can see
that the result for the PT approach is slightly higher than for the
APT one for large $x$-values. The difference $x\Delta q_{3}^{\rm
PT}(x)-x\Delta q_{3}^{\rm APT}(x)$ vs. $x\,$ is more transparently shown on
Fig.~\ref{figdq32}.

\begin{figure}[h]

     \begin{minipage}[b]{0.75\textwidth}%\centering
\vspace*{1.0cm}
\hspace*{3.5cm}
\centering\includegraphics[width=0.55\textwidth]{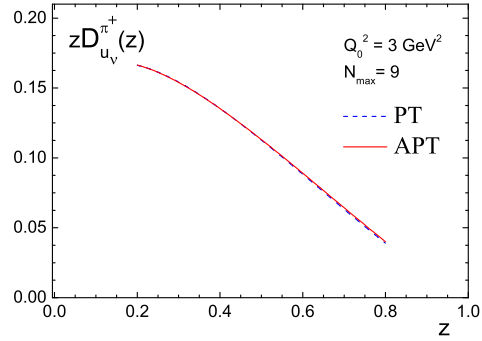} %\centering
    \end{minipage}
\caption {The $D^{\pi^{+}}_{u_v}(z)$-shape obtained in APT (solid
line) and PT (dashed line). }
\label{figD}
   \end{figure}

For the  `fake' data of the nonsinglet combination of the
fragmentation functions  $D^{\pi^{+}}_{u_v}(z,Q^2)$ we have obtained
a very similar shape for PT and APT approaches (see
Fig.~\ref{figD}). The values of the scale parameter are:
$\Lambda_{\rm APT}=307\pm 25$ MeV and $\Lambda_{\rm PT}=231\pm
12$~MeV.

%%%%%%%%%%%%%%%%%%%  FIGURE  1 - 2 %%%%%%%%%%%%%%%%%%%%%%%%%%%%%%

%%%%%%%%%%%%%%%%%%%  FIGURE - 3 %%%%%%%%%%%%%%%%%%%%%%%%%%%%%%
%###########################################################################

In general, for both nonsinglet combinations  $x\Delta q_3(x)$ and
$zD^{\pi^{+}}_{u_v}(z)$ the PT result is higher than for the APT one
for large $x$ or $z$ respectively. The same property we have for
$xF_3(x)$ structure function \cite{SS-F3-2}. We confirm the inequality
$\Lambda_{\rm APT} > \Lambda_{\rm PT}$, obtained previously for
$xF_3(x)$ structure function.

It should be noted that kinematic area for variable $z$ is
considerable narrower than the kinematic region for  variable $x$.
This may be the reason that the behavior of the
$zD^{\pi^{+}}_{u_v}(z,Q^2)$ function in the PT and APT
approximations are practically the same (see Fig.~\ref{figD}).

%#########################################################
\section*{Acknowledgments}
%#########################################################
It is a pleasure for the authors to thank D.V. Shirkov for
stimulating discussions and to A.~E.~Do\-rokhov, V.~L. Khandramai,
S.~V.~Mikhailov, and  O.~V.~Teryaev for interest in this work.

This research was supported by the JINR--BelRFFR grant F14D-007, the
Heisenberg--Landau Program 2014, JINR--Bulgaria Collaborative Grant,
and by the RFBR Grants (Nrs 12-02-00613, 13-02-01005 and
14-01-00647).

%%%%%%%%%%%%%%%%%%%%%%%%%%%%%%%%%%%%%%%%%%%%%%%%%%%%%%%%%%%%%%%%%%%%%%%

\end{document}